\documentclass[runningheads]{llncs}
\usepackage{graphicx}
\usepackage{color}

\definecolor{cb_orange}{rgb}{1.0,0.51,0.0}
\definecolor{cb_blue}{rgb}{0.22,0.49,0.72}
\definecolor{cb_green}{rgb}{0.3,0.67,0.29}
\definecolor{cb_red}{rgb}{0.89,0.1,0.11}
\definecolor{cb_pink}{rgb}{1, 0, 0.4}
\definecolor{cb_purple}{rgb}{0.5,0.0,0.5}
\definecolor{cb_cyan}{rgb}{0.0,1.0,1.0}
\definecolor{cb_magenta}{rgb}{1.0,0.0,1.0}

\usepackage{amssymb} 
\usepackage{algorithm}
\usepackage{algpseudocode}
\usepackage{multirow}
\usepackage{mathrsfs}
\usepackage{amsmath}
\usepackage{graphicx}
\usepackage{color}
\usepackage{upgreek}
\usepackage{bbm}
\usepackage{bm}
\usepackage[utf8]{inputenc}
\usepackage{url}
\usepackage{microtype} 

\begin{document}

\author{Kyungryun Lee \and Won-Ki Jeong\thanks{Corresponding author}}

\authorrunning{K. Lee and W.K. Jeong}

\institute{Korea University, College of Informatics, \\
 Department of Computer Science and Engineering \\
\email{\{krlee0000, wkjeong\}@korea.ac.kr}}
%
%
%


\title{Reference-free Axial Super-resolution of \\3D Microscopy Images using Implicit Neural Representation with a 2D Diffusion Prior}
\titlerunning{3D Microscopy Axial SR using INR with a 2D Diffusion Prior}


\maketitle              
\begin{abstract}
Analysis and visualization of 3D microscopy images pose challenges due to anisotropic axial resolution, demanding volumetric super-resolution along the axial direction.
While training a learning-based 3D super-resolution model seems to be a straightforward solution, it requires ground truth isotropic volumes and suffers from the curse of dimensionality.
Therefore, existing methods utilize 2D neural networks to reconstruct each axial slice, eventually piecing together the entire volume.
However, reconstructing each slice in the pixel domain fails to give consistent reconstruction in all directions leading to misalignment artifacts.
In this work, we present a reconstruction framework based on implicit neural representation (INR), which allows 3D coherency even when optimized by independent axial slices in a batch-wise manner.
Our method optimizes a continuous volumetric representation from low-resolution axial slices, using a 2D diffusion prior trained on high-resolution lateral slices without requiring isotropic volumes.
Through experiments on real and synthetic anisotropic microscopy images, we demonstrate that our method surpasses other state-of-the-art reconstruction methods.
The source code is available on GitHub: \url{https://github.com/hvcl/INR-diffusion}.
\keywords{Implicit neural representation \and Isotropic reconstruction \and Diffusion models}
\end{abstract}

\section{Introduction}
\label{intro}
High-resolution 3D microscopy images are widely used in biomedical research to reveal micro- to nano-scale structures in tissue samples. 
However, their axial resolution is often coarser than their lateral resolution due to physical limitations such as tissue sectioning or light diffraction. 
This substantial difference in resolution, which can be up to an order of magnitude, results in significant visual artifacts when the samples are viewed from the side, making analysis and downstream tasks difficult. 
Therefore, there is a growing need for isotropic reconstruction of 3D microscopy images, also known as axial super-resolution (SR).

Interpolation and deconvolution algorithms are traditional axial reconstruction methods that offer quick solutions without the need for a training process. 
However, they often produce suboptimal, blurry results, particularly for data samples rich in texture.
More recent deep-learning methods can alleviate this issue, such as supervised training of a 3D neural network using anisotropic/isotropic paired 3D training data~\cite{3D-SR-Unet}.
However, training a 3D neural network faces two major challenges: the lack of 3D isotropic training data and the high cost of training a 3D neural network.
Since directly using a 3D neural network is challenging, recent studies utilize 2D neural networks and generative models. 
Weigert et al.~\cite{IsoNet,CSBdeep} proposed training a 2D U-Net\cite{Unet} with the lateral images of the target 3D volume and then reconstructing the axial slices individually.
Deng et al.~\cite{deng} utilized Cycle-GAN to learn the degradation process between axial and lateral images, then applied this knowledge to train a 2D super-resolution model for reconstructing the axial slices. 
More recently, advanced generative models, such as diffusion models~\cite{ddpm}, are actively adopted to improve the axial SR quality, either by reconstructing the 3D volume in a slice-by-slice manner~\cite{diff-em,diffuseIR} or approximating the 3D data prior with the product of two perpendicular 2D diffusion priors~\cite{tpdm}. 
Nevertheless, as the reconstructed slices are not implicitly connected, approximating the volume with individual slices might fail to capture the 3D coherency of the volume and often cause misalignment artifacts.

To overcome the above limitations, we propose a novel axial SR method based on implicit neural representation(INR), which is  
%
a multi-layer perceptron that maps continuous spatial coordinates $(x,y,z)$ to their intensity value. 
INR has shown great success in inverse rendering which allows reconstruction in a continuous volumetric representation from discrete measurements~\cite{nerf} and was later applied to diverse medical imaging reconstruction problems~\cite{nerp,dynamic-CT,single-subject,intratomo}.
Instead of using a 2D diffusion model to directly reconstruct a 3D isotropic volume in the pixel domain, our idea is to optimize the INR using a 2D diffusion model via score distillation sampling to implicitly learn 3D spatial coherency across the volume. 
To evaluate the effectiveness and practicality of our approach, we conducted extensive experiments on various datasets with high anisotropic axial scaling factors ($\times$8, $\times$10). 
Our experiments included FIB25 for simulation studies, CREMI for a real anisotropic electron microscopy image of Drosophila brain tissue, and a real two-channel fluorescent microscopy image of Zebrafish retina cells. 
We compared our method with other state-of-the-art techniques to show its reconstruction quality and effectiveness.
The results demonstrate that our method produces results that are not only detailed and reliable but also accurate to the low-resolution data.

\section{Method}
\label{method}
\begin{figure}[t] 
    \centering
    \includegraphics[width=0.99\textwidth]{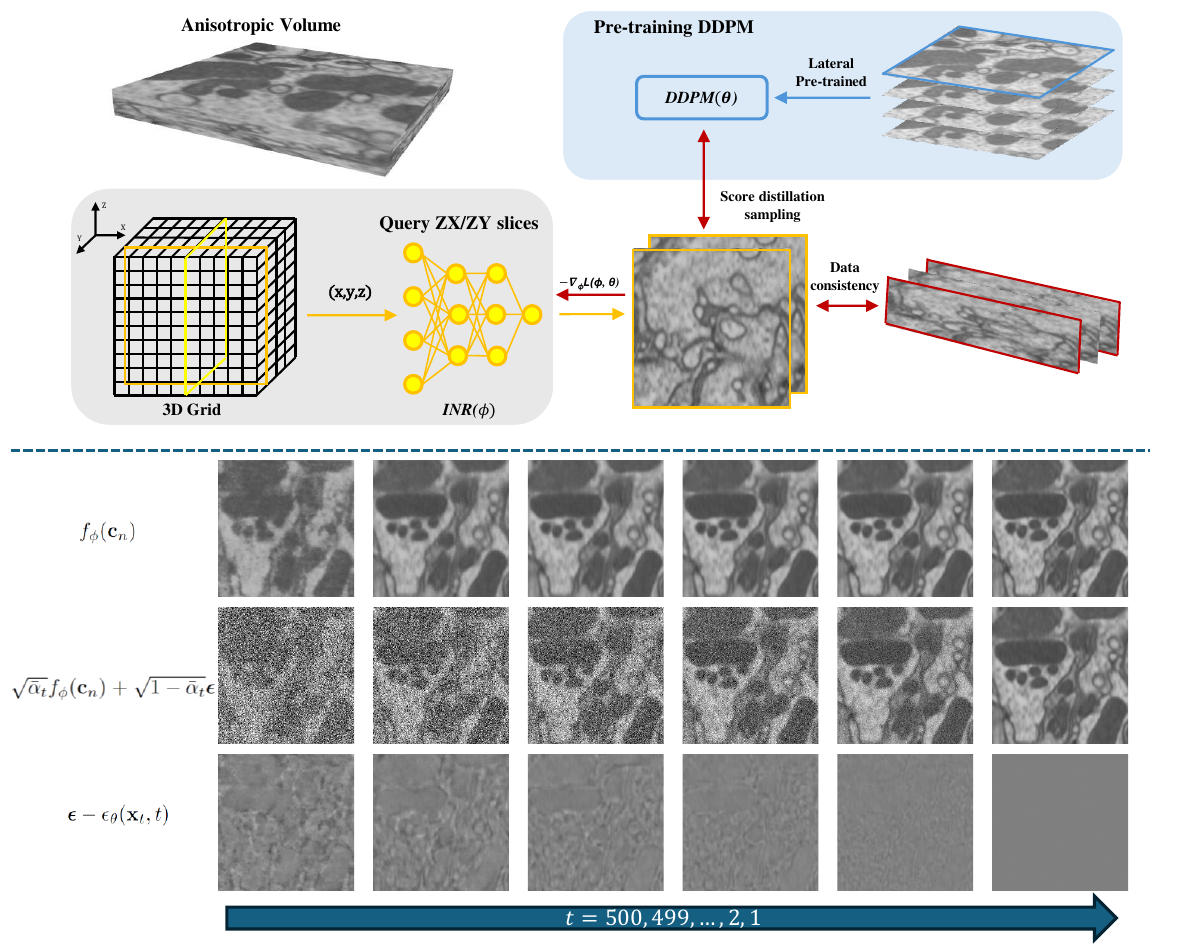}
    \caption{
    (top): The overview of our framework. First, we train the diffusion model using XY slices of the anisotropic volume, where the Z resolution significantly degrades compared to the XY resolution.
    During the optimization process of the INR, we randomly sample batches of ZX,ZY planes and optimize the loss.
    (bottom): The evolution of applying the diffusion prior: queried image(first row), $t$-step noisy image(second row) and the scaled difference between the original image and the refined output of the diffusion model.
    The discrepancy between the image before and after passing the diffusion model provides a guiding direction.
    %
    }
    \label{fig:framework}
\end{figure}

The overview of our proposed model is illustrated in Fig.~\ref{fig:framework}(top).
As 3D anisotropic volume does not exist, we pre-train a 2D diffusion model with high-resolution lateral slices(XY).
Once we have a diffusion model, we train an INR by querying axial slices(ZX or ZY) and fitting it to two loss functions: the measurement consistency loss and a diffusion prior loss.
The measurement consistency loss ensures that the queried image matches the low-resolution measurement after passing the forward degradation process.
By utilizing the score distillation sampling\cite{poole2022dreamfusion,red-diff}, we incorporate the lateral image pre-trained diffusion model as an image prior for sharp and detailed reconstruction.
A more detailed description of each component of our method is provided in the following sections.

\subsection{Implicit neural representations for isotropic volume reconstruction}

Implicit neural representations encode a spatial data field into the weights of a neural network; in other words, the model learns to map spatial coordinates of a pixel$(x, y, z)$ to its data (e.g., intensity) values.
The function is parameterized by $\phi$ and defined as $f_\phi : \textbf{c} \rightarrow  f_\phi(\textbf{c})$, where the input $\textbf{c} \in \mathbb{R}^{d}$ is the $d$ dimension spatial coordinates($d=3$ for $(x,y,z)$) and the output $f_\phi(\textbf{c}) \in \mathbb{R}$ is the scalar intensity value for the given the position. 
The input spatial coordinates are initially projected to a higher-dimensional space, and then processed through a multi-layer perceptron(MLP) architecture, returning an intensity value as an output.
As the coordinates are implicitly connected via a neural network, 
training the MLP can be conducted using separate slice and coordinate pairs. 

Consider reconstructing an isotropic volume in the pixel space where it can be approximated as reconstructing $N$ sequential axial slices(ZX or ZY).
Each of the $N$ 2D reconstruction problem is formulated as 
$\textbf{y}_n = \textbf{A}\textbf{x}_n + \bm{\epsilon}$
where $\textbf{x}_n$ represents the $n$-th unseen isotropic slice, $\textbf{y}_n$ is the $n$-th downsampled slice(measurement) with additive noise $\bm{\epsilon}$ and $\textbf{A}$ is the Z direction degradation matrix shared along all the slices.
One can reconstruct the $n$-th axial slice $\textbf{x}^*_n$ by solving the equation 

\begin{equation}
    \textbf{x}^*_n = \operatorname{argmin}_{\textbf{x}} ||\textbf{Ax}-\textbf{y}_n||^2_2 + R(\textbf{x})
    \label{eq:argmin}
\end{equation}
where the first term ensures data fidelity by maintaining consistency between \textbf{x} and \textbf{y}, while the second term represents image regularization which incorporates prior knowledge of the image.
%
%
Reconstructing the $N$ slices independently and subsequently concatenating 
them together as a volume can lead to misalignments due to the highly ill-posed nature of each reconstruction problem.
Instead, we optimize a single coordinate-based INR to fit the $n$ equations.
This approach enables the INR to implicitly learn 3D spatial coherency.

To train the MLP, we randomly query a $n$-th axial plane $f_\phi(\textbf{c}_n)$ by forwarding its spatial coordinates 
$\textbf{c}_n$   
and minimize a loss that resembles Eq.~\ref{eq:argmin}
\begin{equation}
    \mathcal{L} (\phi) = 
    \mathbb{E}_{n} 
    \left[
    ||\textbf{A}f_\phi(\textbf{c}_n) - \textbf{y}_n||^2_2 + \lambda R(f_\phi(\textbf{c}_n))
    \right]
    \label{eq:INR_loss}
\end{equation} 
where $\lambda$ balances the strength between data fidelity and the regularization loss.
Often choices of $R(\cdot)$ is the total-variation(TV) loss which gives smoother reconstruction. 
However, in this study, we apply a diffusion prior for the regularization term which allows realistic and detailed reconstruction. 
The formulation of this diffusion regularization will be discussed in the next section.

\subsection{Diffusion models and score distillation sampling}

Diffusion models first establish a $T$-step Markovian forward process, gradually perturbing an image with gaussian noise until it approaches to $\textbf{x}_T\sim{\mathcal{N}(\textbf{0},\textbf{I})}$\cite{ddpm,ddpm_vs_gan}. 
The transistion of $\textbf{x}_0$ to $\textbf{x}_t$ can be expressed in a closed form of
\begin{equation}
    q({\textbf{x}_t}|\textbf{x}_0)=\mathcal{N}(\textbf{x}_t;\sqrt{\bar\alpha_t}\textbf{x}_0,(1-\bar\alpha_t)\textbf{I}),
    \label{eq:forward}
\end{equation}
where $\beta_t$, $\alpha_t=1-\beta_t$ and $\bar{\alpha_t}=\prod_{s=1}^{t}\alpha_s$ are the noise scheduling hyper
parameters.
During the reverse process, it generates an image by sampling from an approximated posterior distribution $p_\theta(\textbf{x}_{t-1}|\textbf{x}_t) \approx q(\textbf{x}_{t-1}|\textbf{x}_t, \textbf{x}_0),$
which can be effectively achieved by training a noise predicting model parameterized by $\theta$ that minimizes a (simplified unweighted)loss:
\begin{equation}
    \mathcal{L} (\theta) = \mathbb{E}_{\epsilon,t} 
    \left[
    ||\bm{\epsilon}-\epsilon_\theta( \textbf{x}_t ,t)||_2^2
    \right]
    , \quad \bm{\epsilon}\sim{\mathcal{N}(0,\textbf{I})}
    \label{eq:diffusion_training}
\end{equation}
This formulation represents unconditional diffusion models.

We pre-train an unconditional diffusion model $\theta$ with the lateral high-resolution slices to learn the high-resolution image prior.
With Score Distillation Sampling(SDS)\cite{poole2022dreamfusion}, we can optimize an INR such that the queried images conform to the pre-learned image prior of $\theta$.
We first query a random axial slice $\textbf{x}_n = f_\phi(\textbf{c}_n)$ and sample a Gaussian noise $\bm{\epsilon}\sim{\mathcal{N}(\textbf{0},\textbf{I})}$.
Following Eq.~\ref{eq:forward}, we add the Gaussian noise to generate a $t$-step noisy image $\textbf{x}_t = \sqrt{\bar\alpha_t}\textbf{x} +\sqrt{1-\bar\alpha_t}\bm{\epsilon}$ (subscript $n$ is omitted from here onwards). 
By forwarding $\textbf{x}_t$ through a frozen diffusion model and minimizing the loss function Eq.~\ref{eq:diffusion_training} with respect to $\phi$(INR), 

\begin{equation}
    \nabla_{\phi} \mathcal{L}(\phi, \theta) = 
    \mathbb{E}_{\epsilon,t} 
    \left[
    (\bm{\epsilon}-\epsilon_\theta( \textbf{x}_t ,t)) 
    \frac{\partial \hat{\epsilon}_\theta(\textbf{x}_t, t)}{\partial \textbf{x}_t}  
    \frac{\partial \textbf{x}_t}{\partial \phi}
    \right]
    \approx
    \mathbb{E}_{\epsilon,t} 
    \left[
    (\bm{\epsilon}-\epsilon_\theta( \textbf{x}_t ,t)) 
    \frac{\partial \textbf{x}_t}{\partial \phi}
    \right]
    \label{eq:sds}
\end{equation}
we optimize the MLP so that it follows the diffusion prior. Note that the Jacobian of the diffusion model, $\frac{\partial \hat{\epsilon}_\theta(\textbf{x}_t, t)}{\partial \textbf{x}_t}$, is omitted which was proven to show comparable performance with much lower computational cost.
%
%
%
Intuitively speaking, we are optimizing the INR to minimize the difference between the original queried image and the refined image after passing the diffusion model.
By applying the SDS as a regularization in Eq.~\ref{eq:INR_loss}, we are able to train an INR representing an isotropic volume that accurately fits the measurements while preserving high levels of detail:
\begin{equation}
    \mathcal{L} (\phi, \theta) = 
    \mathbb{E}_{n,\bm{\epsilon},t} 
    \left[
    ||\textbf{A}f_\phi(\textbf{c}_n) - \textbf{y}_n||^2_2 +
    \lambda\text{\texttt{stopgrad}}(\bm{\epsilon}-\epsilon_\theta( \textbf{x}_t ,t))^\intercal
    \\
    f_\phi(\textbf{c}_n)
    \right]
    \label{eq:final_loss}   
\end{equation}
where $\textbf{x}_t = \sqrt{\bar\alpha_t}f_\phi(\textbf{c}_n) +\sqrt{1-\bar\alpha_t}\bm{\epsilon}$ is the $t$-level noisy image and \text{\texttt{stopgrad}} is used to represent the gradient Eq.~\ref{eq:sds} in the form of a loss function.

We linearly decrease the level of noise starting from 500(or 250) as the optimization progresses, $t= 500,499,..,1$.
At the initial steps of the optimization, higher regularization of the diffusion model is applied, while towards the end, only fine details are tuned as shown in Fig.~\ref{fig:framework}(bottom).

\section{Experiments}
\label{experiment}
\subsection{Datasets and implementation details}
We evaluate our framework by conducting a performance evaluation on three volumetric microscopy datasets.
We use FIB-25 which is an isotropic FIB-SEM\cite{fib25} dataset to conduct simulation studies (i.e., making synthetic downsampled data) for quantitative evaluation.
For our real-world data applications, we use CREMI\cite{cremi}, which is a ssTEM data with an axial downsampling factor of 10, and a two-channel fluorescence microscopy image of a zebrafish retina sampled at a downsampling factor of 10.2~\cite{retina,CSBdeep}. 
As there is no isotropic ground-truth data available for those two datasets, we only conducted qualitative assessments on them.
We use linear interpolation as our baseline method.

We compare our framework with state-of-the-art isotropic reconstruction methods which are all based on diffusion models.
We use a same diffusion model for all the methods including ours.
Before starting the reconstruction(or optimization) the shared diffusion model is pre-trained with only 2D lateral images of the target volume where there is no isotropic volume to use as a reference.
The diffusion model is trained using the backbone and setting of ~\cite{ddpm} where $T=1000$ and the learning rate is 0.00002. 
For the methods\cite{diff-em,diffuseIR} that take a slice-by-slice approach we reconstruct ZX slices along the y axis.

Considering that our microscopy images contain lots of high-frequency details, we implement the INR with an MLP consisting of sine activation~\cite{siren} layers and a Gaussian Fourier feature embedding at the beginning~\cite{ffe} to capture the fine details. 
The MLP comprises a fully connected layer with a width of 768 and a depth of 8, along with a 512-dimensional random Gaussian Fourier feature embedding sampled from a standard deviation of 16.
For the CREMI and FIB-25 datasets, we set the strength of the diffusion regularizer as $\lambda=0.25$, with a decreasing level of noise starting from $t=500$ to $t=1$. The MLP is trained with a learning rate of 0.00001 for 500 epochs, allowing the model to observe the volume 500 times. The volume is divided into batches of 8 random axial slices, and the reconstruction direction (ZX or ZY) is alternated with each iteration.

The training protocol remains consistent for the two-channel fluorescence microscopy volume, except that the MLP returns two outputs for each channel and is trained for 200 epochs.

\subsection{Simulation studies on FIB25}\label{subsec3.2}
We randomly extracted a 512×512×512 sub-volume, convolved it with a 3D Gaussian filter stretched in the Z-direction with a standard deviation of 4, and then downsampled it by selecting every 8 lateral slices along the Z-direction to simulate a 64×512×512 anisotropic volume.
The 64 lateral images are used to train the diffusion model, wherein they are randomly cropped to a size of 256×256.
We reconstruct a subpart of the training volume of size 32×256×256 and evaluate ZX, ZY, XY planes with PSNR, Structural similarity index measure(SSIM), and LPIPS~\cite{lpips} to assess human perceptual quality.
\begin{table}[tb]
\centering
\caption{Quantitative Comparison with other methods. With a Gaussian filter ($\sigma_z/\sigma_{xy} = 4/0.5$) and a downsampling rate of 8, the isotropic FIB25 volume is simulated to an anisotropic resolution of $32\times256\times256$. The baseline is linear interpolation. The highest scores are highlighted in \textbf{bold} and the second is underlined.}
\label{table:comp}
\begin{tabular}{|c|ccc|ccc|ccc|}
\hline
\multirow{2}{*}{\textbf{Method}} & \multicolumn{3}{|c|}{\textbf{ZX}}  &\multicolumn{3}{|c|}{\textbf{ZY}} & \multicolumn{3}{|c|}{\textbf{XY}} \\
& PSNR$\uparrow$ & SSIM$\uparrow$ & LPIPS$\downarrow$ & PSNR$\uparrow$ & SSIM$\uparrow$ & LPIPS$\downarrow$ & PSNR$\uparrow$ & SSIM$\uparrow$ & LPIPS$\downarrow$ \\
\cline{1-10}
Baseline & 26.38 & 0.544 & 0.535 & 26.38 & 0.541 & 0.539 & 26.51 & 0.585 & 0.398 
\\
\cline{1-10}
Lee et al.\cite{diff-em} & 28.62 & 0.665 & 0.423 & 28.63 & 0.664 & 0.416 & 28.78 & 0.688 & \textbf{0.278} 
\\
\cline{1-10}
DiffuseIR\cite{diffuseIR} & 26.16 & 0.524 & 0.456 & 26.16 & 0.517 & 0.527 & 26.34 & 0.548 & 0.411 
\\
\cline{1-10}
TPDM\cite{tpdm} & 28.57 & 0.663 & \textbf{0.357} & 28.58 & 0.661 & \textbf{0.356} & 28.65 & 0.687 & \underline{0.296} 
\\
\cline{1-10}
Ours & \textbf{28.78} & \textbf{0.667} & \underline{0.410} & \textbf{28.78} & \underline{0.664} & \underline{0.406} & \textbf{28.88} & \textbf{0.688} & 0.359 
\\
\hline
\end{tabular}
\end{table}

\begin{figure}[t]
    \centering
    \includegraphics[width=0.99\textwidth]{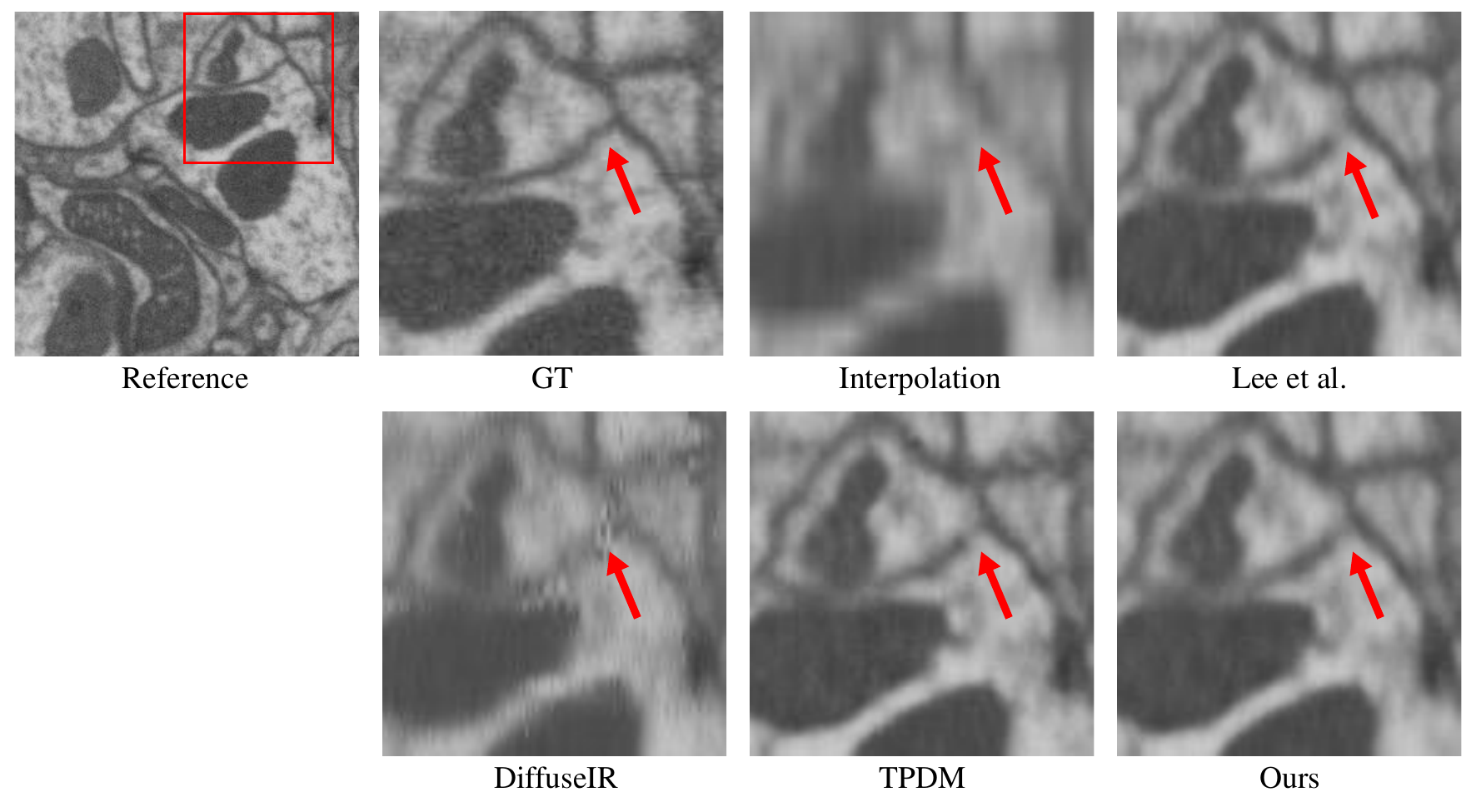}
    \caption{
    Visual comparison of reconstructing the simulated FIB25 volume with a downsampling factor of 8. The volume is viewed in the ZY direction. 
    All methods except ours show disconnection in the intersection.
    }
    \label{figure:fib25}
\end{figure}

Table~\ref{table:comp} shows that our framework offers the highest PSNR/SSIM score.
TPDM~\cite{tpdm}, which is a diffusion model based SOTA 3D reconstruction method originally used for MRI Z-axis SR, shows the highest LPIPS score showing texture-rich reconstruction, however it falls behind in PSNR/SSIM.
This indicates that our method produces more reliable results with less hallucination which is common in generative models.
Fig.\ref{figure:fib25} also demonstrates that our method yields more reasonable results for uncertain regions. 
This can be attributed to the implicit 3D connectivity of the INR. 
During the alternating optimization with ZX and ZY slices, information exchange occurs, complementing the uncertainty of each region.
Moreover, the implicit 3D coherency prevents misalignments, unlike DiffuseIR, which suffers from severe artifacts due to separate reconstruction of ZX slices.
By indirectly applying the diffusion model, our method is capable of capturing both high-frequency details and structural information while preserving data consistency.

\subsection{Real world anisotropic volumes}
\begin{figure}[t]
    \centering
    \includegraphics[width=0.99\textwidth]{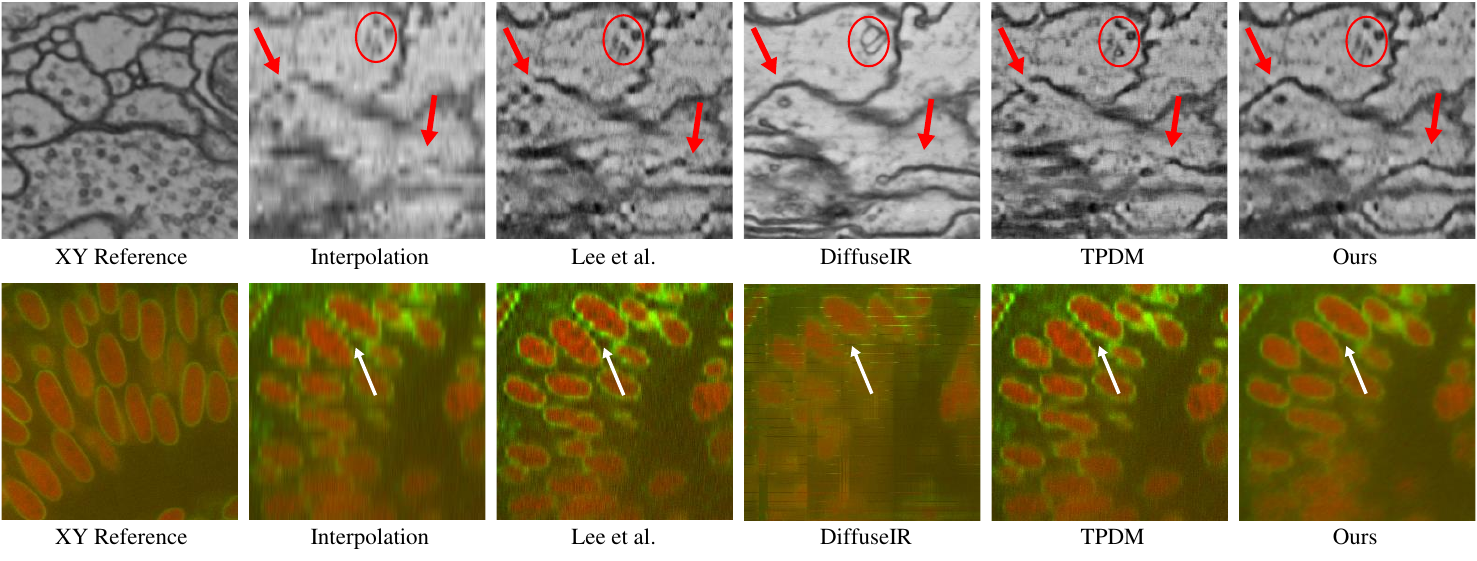}
    \caption{
    (row1): Visual comparsion of reconstructing real anisotropic electron microscopy image with a downsampling factor of 10. Inside the red circle shows an example of hallucination. As there is no ground truth image, we use the lateral slices(XY) as a reference. 
    (row2): A merged fluorescence microscopy image of a zebrafish retina cell composed of two channels: the red channel representing the nuclei and the green channel representing the nuclear envelope.
    }
    \label{figure:CREMI_Retina}
\end{figure}

We reconstruct two real anisotropic microscopy images to demonstrate the practical usage of our method. 
Both datasets have dimensions of 26×256×256, and our goal is to restore them to dimensions of 256×256×256. 
Since we do not have prior knowledge of the degradation process, we assume linear downsampling for all methods.

In the CREMI data, severe downsampling rates lead to disconnections between horizontal structures. 
While other methods fail to restore these disconnections, our method successfully addresses this issue. 
As shown in Fig.\ref{figure:CREMI_Retina}(row1), boundary disconnection is commonly shown in other methods, whereas in our case they are smoothly connected. 
Furthermore, while hallucinations may occur when the diffusion model directly generates the reconstructed result, our method utilizes the diffusion model only as a regularizer indirectly, thereby avoiding such cases.

The zerbrafish retina image consists of two channels where the nuclei and nuclear envelope is seperatley imaged.
In such situations, the INR shows strong adaptability because it can easily share information between different channels. 
By leveraging the correlation between the channels, it allows not only proper channel alignment but also mutual complementation. 
The second row of the figure proves that our approach effectively eliminates the disconnection artifacts on the nuclear envelope and accurately aligns it with the inner side.

\section{Conclusion and Future work}
\label{conclusion}
We introduce a microscopy axial super-resolution framework that utilizes 2D diffusion models, eliminating the need for an isotropic 3D volume.
By employing score distillation to enforce a 2D diffusion prior on an implicit neural representation, we can efficiently reconstruct the volume without the lose of 3D coherency.
Through evaluation on three datasets, we demonstrate the superior reconstruction capabilities of our framework compared to other state-of-the-art diffusion model based methods.
Results show that our method not only generates detailed reconstructions but also ensures reliability with reduced risk of hallucination.

While our method utilizes SDS, which does not require back-propagation through the diffusion model, optimizing the INR involves fitting the volume hundreds of times, resulting in a bottleneck in training time.
Future research will focus on achieving faster optimization by adapting advanced techniques for both the diffusion model and the INR.

\subsubsection{Acknowledgements.}
\label{acknowledgements}
This work was partially supported by the National Research Foundation of Korea (RS-2024-00349697, NRF-2021R1A6A1A13044830), the Institute for Information \& Communications Technology Planning \& Evaluation (IITP-2024-2020-0-01819), the Technology development Program(RS-2024-00437796) funded by the Ministry of SMEs and Startups(MSS, Korea), and a Korea University Grant.

\subsubsection{Disclosure of Interests.}
There are no conflicts of interest to declare.

\bibliographystyle{splncs04}
\bibliography{reference}

\begin{thebibliography}{10}
\providecommand{\url}[1]{\texttt{#1}}
\providecommand{\urlprefix}{URL }
\providecommand{\doi}[1]{https://doi.org/#1}

\bibitem{deng}
Deng, S., Fu, X., Xiong, Z., Chen, C., Liu, D., Chen, X., Ling, Q., Wu, F.: Isotropic reconstruction of 3d em images with unsupervised degradation learning. In: Medical Image Computing and Computer Assisted Intervention--MICCAI 2020: 23rd International Conference, Lima, Peru, October 4--8, 2020, Proceedings, Part V 23. pp. 163--173. Springer (2020)

\bibitem{ddpm_vs_gan}
Dhariwal, P., Nichol, A.: Diffusion models beat gans on image synthesis. Advances in Neural Information Processing Systems  \textbf{34},  8780--8794 (2021)

\bibitem{cremi}
Funke, J., Saalfeld, S., Bock, D., Turaga, S., Perlman, E.: Miccai challenge on circuit reconstruction from electron microscopy images, \url{https://cremi.org/}

\bibitem{3D-SR-Unet}
Heinrich, L., Bogovic, J.A., Saalfeld, S.: Deep learning for isotropic super-resolution from non-isotropic 3d electron microscopy. In: Medical Image Computing and Computer-Assisted Intervention- MICCAI 2017: 20th International Conference, Quebec City, QC, Canada, September 11-13, 2017, Proceedings, Part II 20. pp. 135--143. Springer (2017)

\bibitem{ddpm}
Ho, J., Jain, A., Abbeel, P.: Denoising diffusion probabilistic models. Advances in neural information processing systems  \textbf{33},  6840--6851 (2020)

\bibitem{retina}
Icha, J., Kunath, C., Rocha-Martins, M., Norden, C.: Independent modes of ganglion cell translocation ensure correct lamination of the zebrafish retina. Journal of Cell Biology  \textbf{215}(2),  259--275 (2016)

\bibitem{diff-em}
Lee, K., Jeong, W.K.: Reference-free isotropic 3d em reconstruction using diffusion models. In: International Conference on Medical Image Computing and Computer-Assisted Intervention Workshops. pp. 235--245. Springer (2023)

\bibitem{tpdm}
Lee, S., Chung, H., Park, M., Park, J., Ryu, W.S., Ye, J.C.: Improving 3d imaging with pre-trained perpendicular 2d diffusion models. arXiv preprint arXiv:2303.08440  (2023)

\bibitem{red-diff}
Mardani, M., Song, J., Kautz, J., Vahdat, A.: A variational perspective on solving inverse problems with diffusion models. arXiv preprint arXiv:2305.04391  (2023)

\bibitem{single-subject}
McGinnis, J., Shit, S., Li, H.B., Sideri-Lampretsa, V., Graf, R., Dannecker, M., Pan, J., Stolt-Ans{\'o}, N., M{\"u}hlau, M., Kirschke, J.S., et~al.: Single-subject multi-contrast mri super-resolution via implicit neural representations. In: International Conference on Medical Image Computing and Computer-Assisted Intervention. pp. 173--183. Springer (2023)

\bibitem{nerf}
Mildenhall, B., Srinivasan, P.P., Tancik, M., Barron, J.T., Ramamoorthi, R., Ng, R.: Nerf: Representing scenes as neural radiance fields for view synthesis. Communications of the ACM  \textbf{65}(1),  99--106 (2021)

\bibitem{diffuseIR}
Pan, M., Gan, Y., Zhou, F., Liu, J., Zhang, Y., Wang, A., Zhang, S., Li, D.: Diffuseir: Diffusion models for isotropic reconstruction of 3d microscopic images. In: International Conference on Medical Image Computing and Computer-Assisted Intervention. pp. 323--332. Springer (2023)

\bibitem{poole2022dreamfusion}
Poole, B., Jain, A., Barron, J.T., Mildenhall, B.: Dreamfusion: Text-to-3d using 2d diffusion. In: The Eleventh International Conference on Learning Representations (2022)

\bibitem{dynamic-CT}
Reed, A.W., Kim, H., Anirudh, R., Mohan, K.A., Champley, K., Kang, J., Jayasuriya, S.: Dynamic ct reconstruction from limited views with implicit neural representations and parametric motion fields. In: Proceedings of the IEEE/CVF International Conference on Computer Vision. pp. 2258--2268 (2021)

\bibitem{Unet}
Ronneberger, O., Fischer, P., Brox, T.: U-net: Convolutional networks for biomedical image segmentation. In: Medical Image Computing and Computer-Assisted Intervention--MICCAI 2015: 18th International Conference, Munich, Germany, October 5-9, 2015, Proceedings, Part III 18. pp. 234--241. Springer (2015)

\bibitem{nerp}
Shen, L., Pauly, J., Xing, L.: Nerp: implicit neural representation learning with prior embedding for sparsely sampled image reconstruction. IEEE Transactions on Neural Networks and Learning Systems  (2022)

\bibitem{siren}
Sitzmann, V., Martel, J., Bergman, A., Lindell, D., Wetzstein, G.: Implicit neural representations with periodic activation functions. Advances in neural information processing systems  \textbf{33},  7462--7473 (2020)

\bibitem{fib25}
Takemura, S.y., Xu, C.S., Lu, Z., Rivlin, P.K., Parag, T., Olbris, D.J., Plaza, S., Zhao, T., Katz, W.T., Umayam, L., et~al.: Synaptic circuits and their variations within different columns in the visual system of drosophila. Proceedings of the National Academy of Sciences  \textbf{112}(44),  13711--13716 (2015)

\bibitem{ffe}
Tancik, M., Srinivasan, P., Mildenhall, B., Fridovich-Keil, S., Raghavan, N., Singhal, U., Ramamoorthi, R., Barron, J., Ng, R.: Fourier features let networks learn high frequency functions in low dimensional domains. Advances in Neural Information Processing Systems  \textbf{33},  7537--7547 (2020)

\bibitem{IsoNet}
Weigert, M., Royer, L., Jug, F., Myers, G.: Isotropic reconstruction of 3d fluorescence microscopy images using convolutional neural networks. In: International Conference on Medical Image Computing and Computer-Assisted Intervention. pp. 126--134. Springer (2017)

\bibitem{CSBdeep}
Weigert, M., Schmidt, U., Boothe, T., M{\"u}ller, A., Dibrov, A., Jain, A., Wilhelm, B., Schmidt, D., Broaddus, C., Culley, S., et~al.: Content-aware image restoration: pushing the limits of fluorescence microscopy. Nature methods  \textbf{15}(12),  1090--1097 (2018)

\bibitem{intratomo}
Zang, G., Idoughi, R., Li, R., Wonka, P., Heidrich, W.: Intratomo: self-supervised learning-based tomography via sinogram synthesis and prediction. In: Proceedings of the IEEE/CVF International Conference on Computer Vision. pp. 1960--1970 (2021)

\bibitem{lpips}
Zhang, R., Isola, P., Efros, A.A., Shechtman, E., Wang, O.: The unreasonable effectiveness of deep features as a perceptual metric. In: Proceedings of the IEEE conference on computer vision and pattern recognition. pp. 586--595 (2018)

\end{thebibliography}

\end{document}